\documentclass[runningheads]{llncs}
\usepackage{graphicx}
\usepackage{sidecap}
\sidecaptionvpos{figure}{t}

\usepackage{multirow, makecell}
\usepackage{graphicx}
\usepackage{booktabs}
\usepackage{hyperref}

\usepackage{subfig}

\usepackage[ruled,vlined]{algorithm2e}
\usepackage{amsmath}

\usepackage{subfiles}

\usepackage{cite}

\hypersetup{
      allcolors=blue,
    colorlinks=true,
    citecolor=blue,
    urlcolor=magenta,
}

\begin{document}
\title{A Lightweight CNN and\\ Joint Shape-Joint Space ($JS^2$) Descriptor for Radiological Osteoarthritis Detection}

\titlerunning{$JS^2$ Descriptor}
%
\author{Neslihan Bayramoglu
\inst{1}\and
Miika T. Nieminen\inst{1,2,3}\and
Simo Saarakkala\inst{1,2,3}}
\authorrunning{N. Bayramoglu et al.}
\institute{
Research Unit of Medical Imaging, Physics and Technology, University of Oulu, Finland \and
Department of Diagnostic Radiology, Oulu University Hospital, Oulu, Finland \and
Medical Research Center, University of Oulu and Oulu University Hospital, Oulu, Finland
}
\maketitle              
\begin{abstract}
Knee osteoarthritis (OA) is very common progressive and degenerative musculoskeletal disease worldwide creates a heavy burden on patients with {reduced quality of life} and also on society due to financial impact.
Therefore, any attempt to reduce the burden of the disease could help both patients and society.
In this study, we propose a fully automated novel method, based on combination of joint shape and convolutional neural network (CNN) based bone texture features, to distinguish between the
knee radiographs with and without radiographic osteoarthritis.   
Moreover, we report the first attempt at describing the bone texture using CNN. Knee radiographs from Osteoarthritis Initiative (OAI) and  Multicenter  Osteoarthritis  (MOST) studies were used in the experiments.
Our models were trained on 8953 knee radiographs from OAI and evaluated on 3445 knee radiographs from MOST.
Our results demonstrate that fusing the proposed shape and texture parameters achieves the state-of-the art performance in radiographic OA detection yielding area under the ROC curve (AUC) of $95.21\%$.

\end{abstract}

\section{Introduction}

Osteoarthritis (OA) is one of the leading cause of disability around the world.
The condition causes pain, stiffness, and limitations in movement.
Several different risk factors have been identified for OA including age, obesity, injury, repetitive use of joints, bone density, muscle weakness, and gender \cite{bijlsma2011osteoarthritis}. The incidence of OA is steeply rising because of the ageing population \cite{bijlsma2011osteoarthritis}.
OA, which is a disease of the developed world, creates a heavy burden on patients with reduced quality of life and also on society due to treatment costs and time off at work.

Although there is no cure for OA, its symptoms can be controlled by modifying the risk factors and with other conservative measures.  
Therefore, early identification and treatment of OA may be critical for decreasing the chronic pain and improve joint function.
It is also important to have a better understanding of the disease to develop new treatments to slow its progression and delay joint replacement surgery \cite{altman2010early}.

\begin{SCfigure}
\includegraphics[height=7cm]{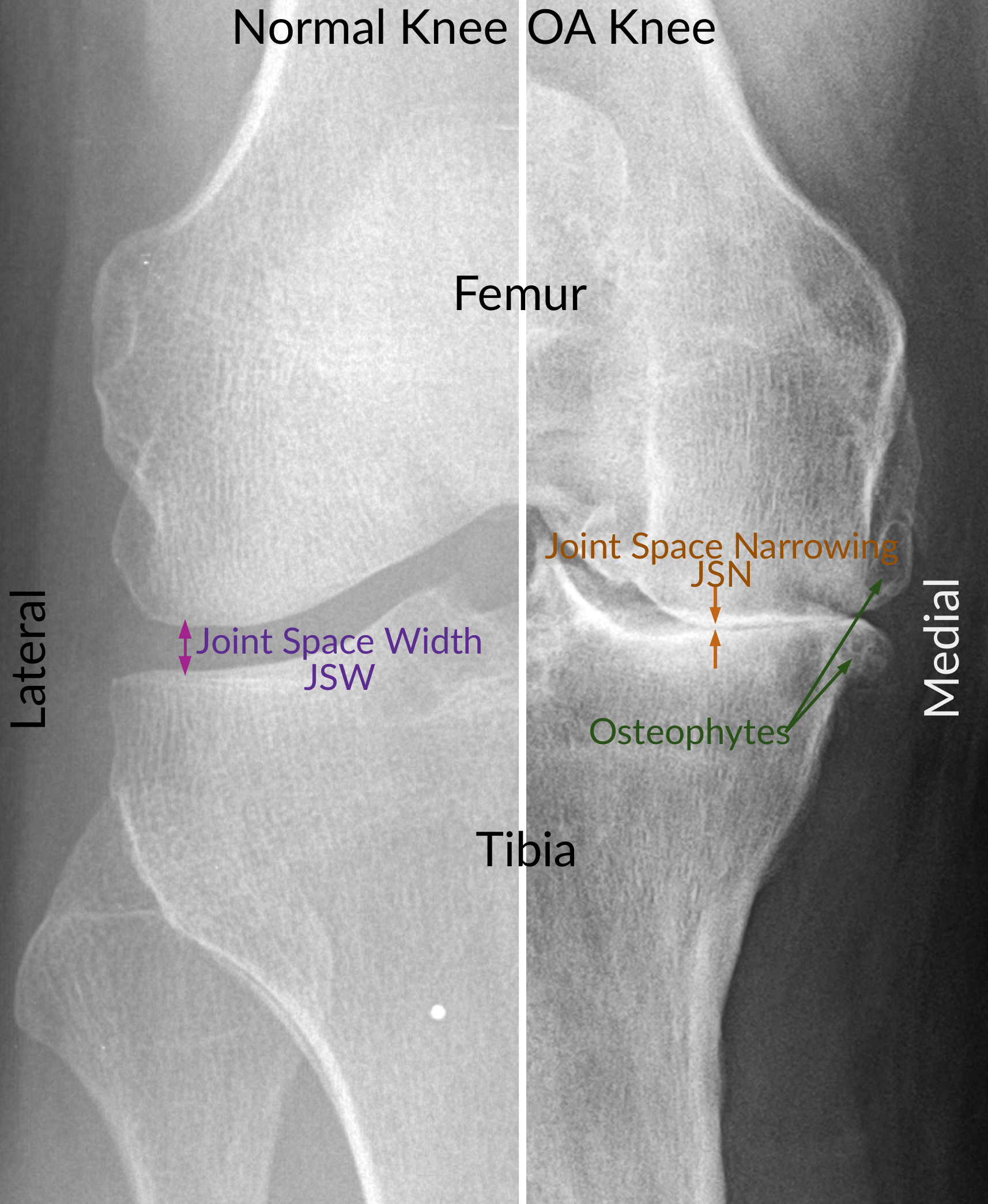}
\caption{\protect\rule{0ex}{8ex}Examples of X-ray images of normal knee and severe OA knee. The left side shows normal knee and the right side shows OA knee.
In this image, in addition to the joint compartments, joint space narrowing (JSN) and osteophytes are demonstrated.}
\label{fig:knee}
\end{SCfigure}

OA is currently diagnosed by clinical examination and, when necessary, with radiography (X-ray imaging) and other advanced imaging modalities such as MRI scanning and arthroscopy. 
Although conventional radiography (X-ray imaging) has several limitations \cite{heidari2011knee}, it is still the primary choice and the most widely used imaging modality in OA because it is more affordable and accessible than other imaging modalities \cite{Haq377}.
X-Ray imaging allows for detection of characteristic features of OA including osteophyte formation, joint space narrowing, subchondral sclerosis, and cysts formation (Figure \ref{fig:knee}).

\begin{figure*}[!htb]
\centering
\includegraphics[width=1\textwidth]{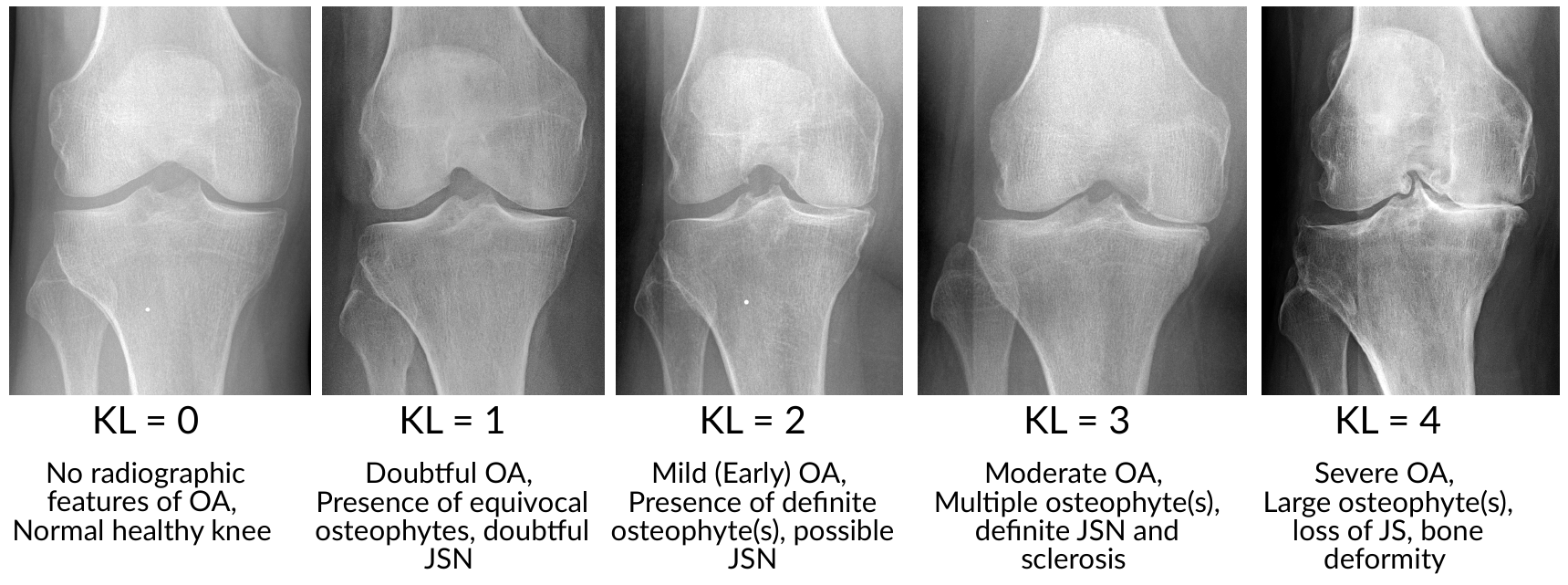}
\caption{Examples of knee X-Rays showing the stages of knee osteoarthritis severity according to KL grading. For the OA detection problem, $KL<2$ is defined as non-OA and $KL>=2$ is classified as OA.}
\label{fig:kl}
\end{figure*}

Several grading systems have been developed to evaluate the OA severity from knee radiographs in clinical practice; semi-quantitative Kellgren and Lawrence (KL) grading system being the most widely used \cite{kellgren1957radiological}. 
In KL grading system, the grade varies from 0 to 4 correlating to increasing severity of OA. 
Figure \ref{fig:kl} demonstrates disease stages according to the KL grading system.
While KL Grade 0 means no radiographic features of OA are present and it is normal,
KL Grade 1 indicates presence of equivocal osteophyte.
KL Grade 2 signifies presence of definite osteophyte and possible joint space narrowing (JSN). 
KL Grade 3 indicates presence of joint space narrowing,  multiple osteophytes, and sclerosis.
KL Grade 4 denotes complete loss of joint space, large osteophytes, severe sclerosis, and definitely bony deformity. International diagnostic threshold for radiographic OA is KL Grade 2:
$KL<2$ is defined as non-OA and $KL>=2$ is classified as OA. It is notable that KL grading suffers from significant inter and intraobserver variation \cite{kose2018inter}.
This is primarily due to subjective differences in interpretations of the X-ray features by observers and different study techniques.

It has been shown in several studies that OA severity assessment from radiographs can be enhanced using computer assisted
methods \cite{lynch1991analysis, kraus2018predictive, hirvasniemi2014quantification, thomson2015automated, janvier2017subchondral, antony2016quantifying, norman2019applying, tiulpin2018automatic, gorriz2019assessing, abedin2019predicting}.
Bone texture analysis is one of the main methodology in the automated analysis of knee radiographs \cite{lynch1991analysis,thomson2015automated,hirvasniemi2014quantification, janvier2017subchondral}.
Bone shape analysis is another popular approach for OA detection \cite{thomson2015automated, minciullo2016fully, haverkamp2011variation, martinez2019discovering}.
However, current {state-of-the-art} (SOTA)
approaches for automatic OA detection and grading are based on machine learning techniques such as deep learning (DL) \cite{antony2016quantifying, norman2019applying, tiulpin2018automatic} and in particular deep convolutional neural networks (CNNs).
The use of deep learning in OA research has been greatly increased in the last few years with the availability of large amounts of annotated data and computational processing power.
However, these SOTA deep learning models suffer from significant shortcomings.
Firstly, well known anatomical features associated with OA were not considered at all.
For example, direct measurement of the separation between the distal femur and the proximal tibia, joint space width \textbf{(JSW)}, 
which is the standard tool for the assessment of knee OA progression is not included \cite{ahlback1968osteoarthrosis}.
Analysis of other radiographic hallmarks of primary OA including joint shape are also missing in these studies.
{Although deep features might cover some aspects of these anatomical features, the added value of explicit inclusion to the models have not been investigated.}
Moreover, in contrast to the common assumption, it was shown that CNNs rely extremely on textures and they perform poorly at recognising shapes without object texture \cite{geirhos2018imagenet}.
Therefore, it is also important to study the outcomes when the bone shape representations are explicitly included.
Second, existing models have huge number of parameters and compute redundant features due to blind region selection.
However, it was found that in OA radiographically most distinctive bony changes occur at the medial tibia margin \cite{bayramoglu2019adaptive}. 
It was observed in \cite{bayramoglu2019adaptive} that this region of interest (ROI) has a profound effect on bone texture analysis.
Existing methods usually resize whole knee joint images blindly for using well known CNN architectures (VGG, ResNet family, etc.) and take the advantage of transfer learning or fine tuning. 
These approaches comes at a price: computational power and substantial amount of labeled data for training.
Lastly, such models are characteristically opaque, i.e., what information in the input data makes them to make their decisions is unclear.

\begin{figure*}[!t]
\centering
\includegraphics[width=1\textwidth]{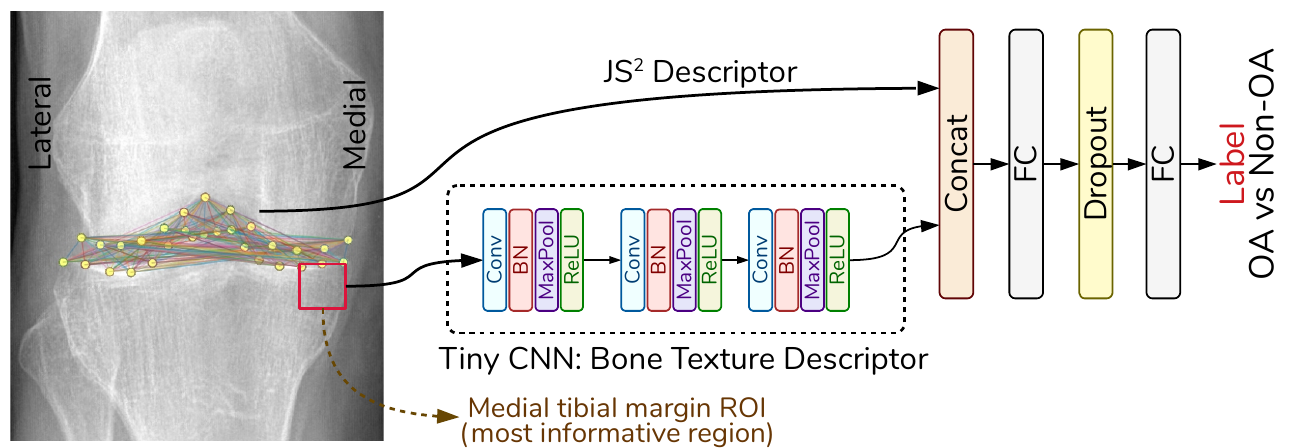}
\caption{We propose a model that combines bone shape and bone texture using a tiny CNN model to detect OA. Our  novel joint  shape  and  joint  space  $JS^2$ descriptor and lightweight CNN features complement each other.}
\label{fig:proposal}
\end{figure*}

In this study, we propose a model that combines bone shape and bone texture using a tiny CNN model to detect OA as defined by $KL>=2$.
Firstly, we propose a compact joint shape and joint space ($JS^2$) descriptor that can be used as a marker of radiographic OA.
Second, we explored whether deep texture features from the tibial margin in the medial side of the knee (the most informative region based on a recent study \cite{bayramoglu2019adaptive}) provides comparable results with the SOTA approaches in which the whole knee joint radiographs were analyzed.
Employing a small ROI enables us to use tiny CNN models that are having significantly less parameters compared to current SOTA DL models and are efficient in terms of both data and computational complexity. Finally, we propose a novel model where both bone texture and joint shape description are combined.

\begin{figure*}[!t]
\centerline{
\subfloat[minJSW and fJSW]{\includegraphics[height=6cm]{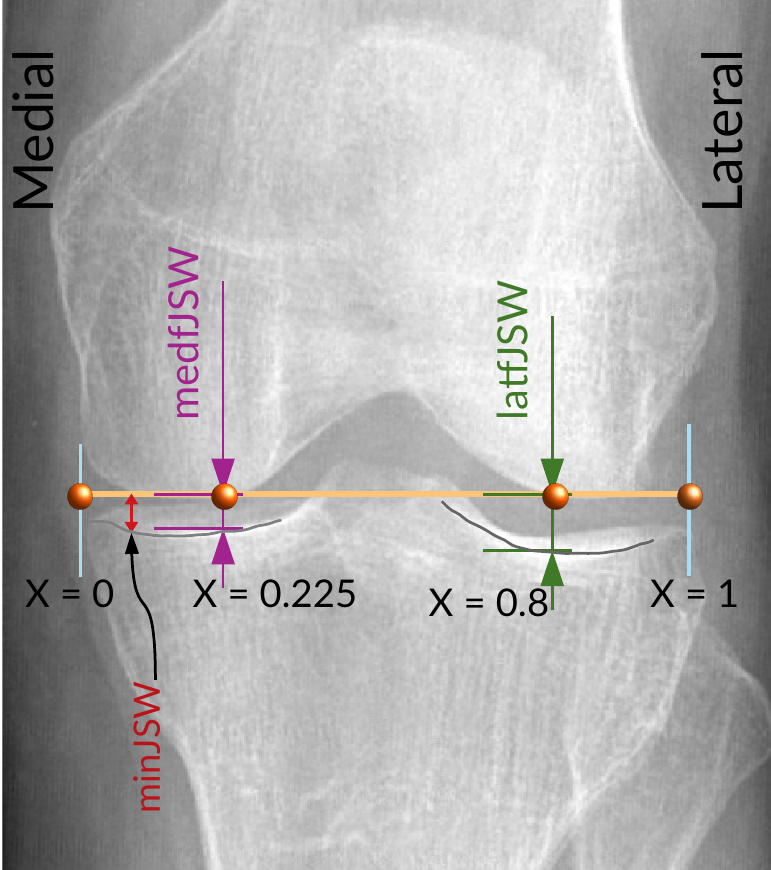}}
\hspace{0.1cm}
\subfloat[$JS^2$]{\scalebox{-1}[1]{\includegraphics[trim=4cm 4cm 4cm 4cm, clip=true,height=6cm]{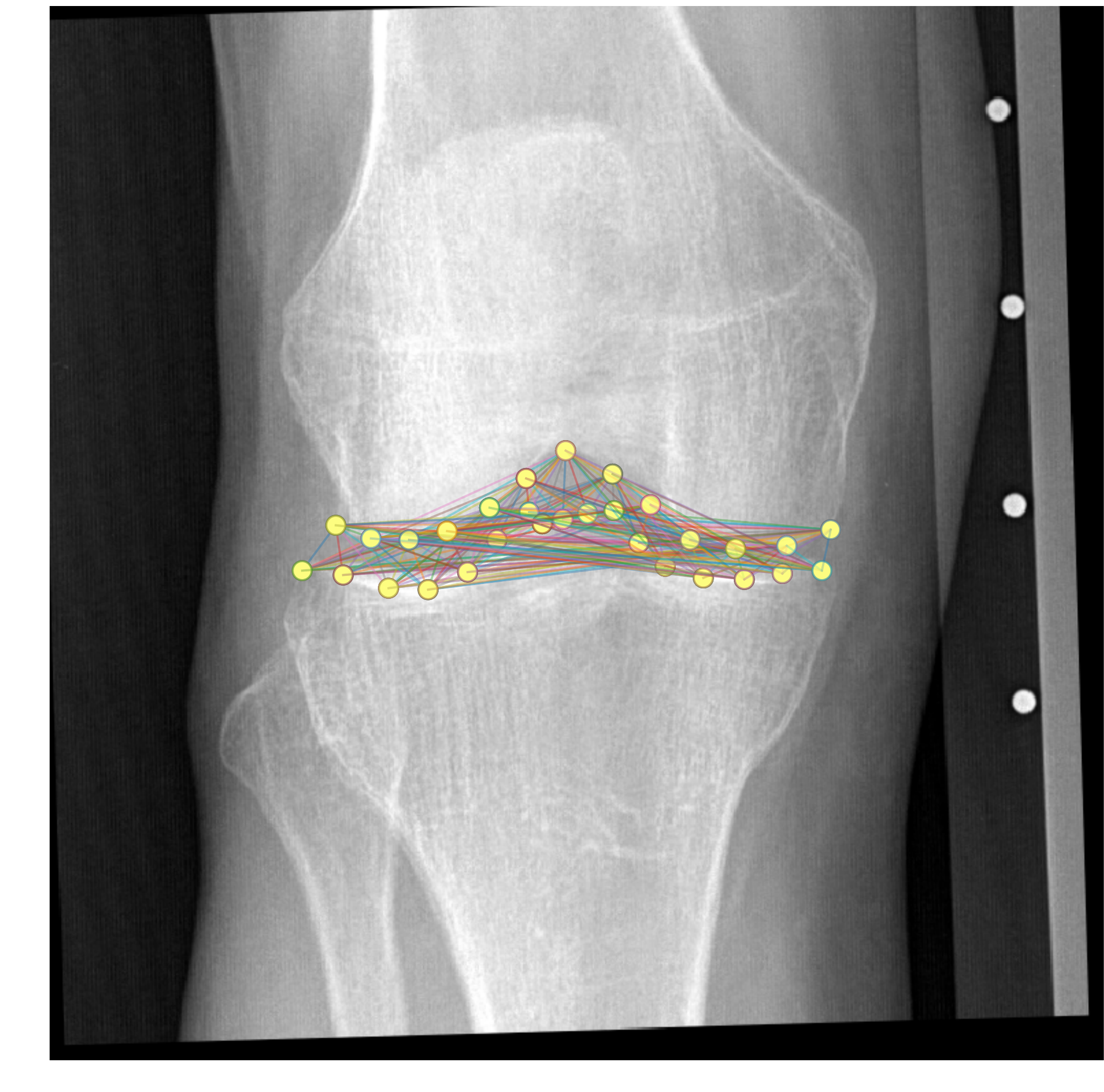}}}
}
\centerline{
{\subfloat[$JS^2$ Connections]{{\includegraphics[trim=1.5cm 0cm 0.5cm 0cm,clip=true,width=0.7\textwidth]{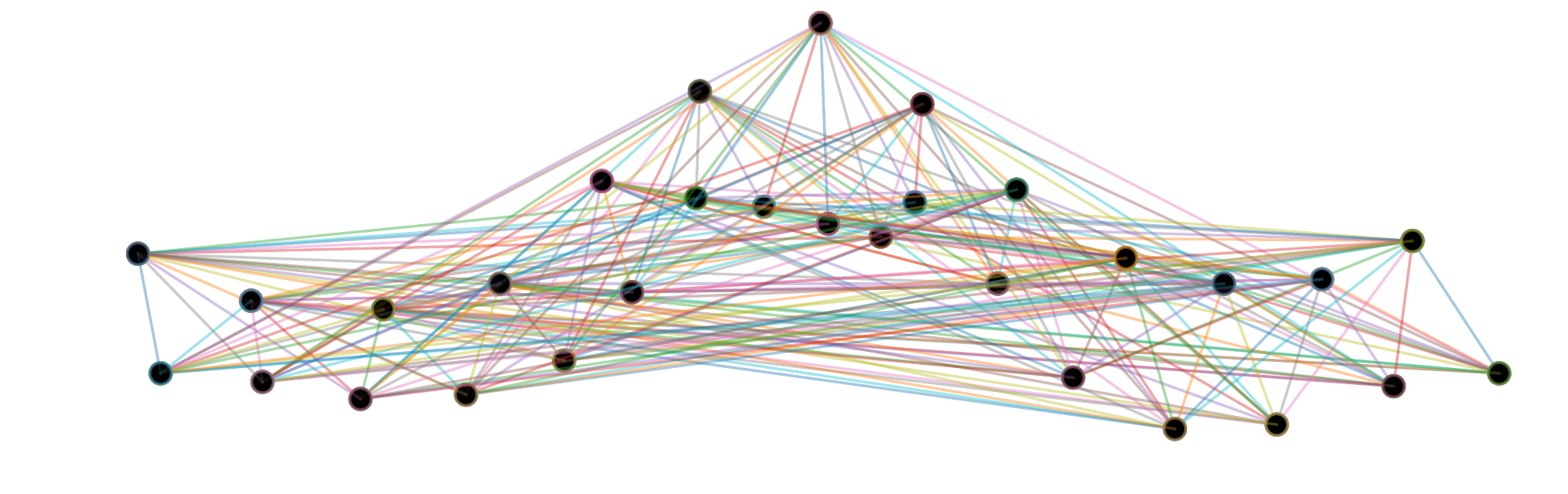}}}}
}
\caption{(a)Illustration of joint space width (JSW) measurements at minimum (minJSW) and at fixed locations (x = 0.225 (medfJSW), x= 0.8 (latfJSW)).
(b) Graphical illustration of our $JS^2$ descriptor on a sample knee radiograph.
(c) $JS^2$ nodes and connections.}
\label{fig:jsw}
\end{figure*}

\section{ Method}

\subsection{Joint Shape and Joint Space ($JS^2$) Descriptor}
With the progression of OA, subchondral bone experiences changes in its structure and composition \cite{buckland2004subchondral,kamibayashi1995trabecular,goldring2009role}.
Previously, it has been shown that the global shape of the tibiofemoral joint is associated with the structural severity of OA \cite{thomson2015automated, haverkamp2011variation}. Joint shape descriptors based on statistical shape modeling have also been successfully employed for radiographic OA detection \cite{minciullo2016fully, thomson2015automated}.

In addition to global shape descriptors, loss of JSW measured between the adjacent bones (femur and tibia) of the knee, which is called Joint Space Narrowing (\textbf{JSN}), has been applied for longitudinal measurement of disease progression\cite{mehta2017comparison}. 
KL grading system is also incorporating assessment of joint space narrowing in addition to the osteophytes.
Change in JSW has been considered a good measure of cartilage volume and thickness as well \cite{neumann2009location}. 
Therefore, minimum joint space width (\textbf{minJSW}) between the projected femur and tibia margins is the currently accepted metric to assess OA severity \cite{neumann2009location, ahlback1968osteoarthrosis}.
However, the reproducibility of radiographic joint space measurements in longitudinal OA studies is problematic and not consistent \cite{guermazi2009plain}.
Although previous studies have improved the reproducibility of minJSW by utilizing location-specific measurement of JSW (\textbf{fJSW}) \cite{neumann2009location, duryea2003new}, JSW measurements do not necessarily translate into a change in radiographic grading of OA \cite{guermazi2009plain}.

\begin{algorithm}[!t]
\SetAlgoLined
 Tibia landmark points $T_{pts}={t_0,t_1,t_2,...,t_{16}}$\;
 Femur landmark points $F_{pts}={f_0,f_1,f_2,...,f_{12}}$\;
 $k\leftarrow 0$\;
 
    \For{$t_i$ in $T_{pts}$}{
      \For{$f_i$ in $F_{pts}$}{
         
            $JS^2_k= \lVert t_i-f_i \rVert $\;
            
            $k \leftarrow k+1$\;
            
            {}
      }
   }
 \caption{$JS^2$ descriptor}
 \label{alg}
\end{algorithm}

In this study, we propose a simple yet efficient descriptor to represent both shape of the knee joint and the spacing of tibiofemoral joint.
Figure \ref{fig:jsw} and Algorithm \ref{alg} represent our proposed $JS^2$ descriptor.
Our method is based on the anatomical landmarks in knee X-ray images and utilizes their spatial relations. 
Here, we utilized BoneFinder\textsuperscript{\textregistered} software \cite{lindner2013fully} in order to locate the landmark points along the contours of the distal femur and the proximal tibia (Figure \ref{fig:jsw_imp}).
It uses random forest regression voting with constrained local model approach.
Although, the BoneFinder\textsuperscript{\textregistered} tool automatically locate 74 points, our $JS^2$ descriptor uses a total of 30 points (13 points on femur contour and 17 points at tibia contour) resulting a vector of length 221($13\times17$) (see Figure \ref{fig:jsw}) .

While previously adopted minJSW and fJSW \cite{neumann2009location, duryea2003new} measurements are used to assess JSN, our $JS^2$ descriptor translates joint shape, femur-tibia configuration and multiple JSW measurements together into a compact descriptor. Therefore, it can be directly utilized to detect radiographic OA.

\subsection{Bone Texture Characterization}

Recent study has considered OA as a whole-organ disease and suggest refocusing from solely cartilage-based studies to bone and other soft tissues together with the cartilage \cite{aspden2019osteoarthritis}.
Since structural changes in subchondral bone occurs many months before changes in articular cartilage thickness (i.e. joint space narrowing) \cite{buckland2004subchondral}, there is a compelling interest towards detecting osteoarthritis with texture analysis of bone
\cite{kraus2018predictive,thomson2015automated}.

Previous studies have reported significant differences in subchondral bone texture between controls and individuals with OA \cite{bayramoglu2019adaptive, mackay2017subchondral, hirvasniemi2014quantification, janvier2017subchondral, kraus2018predictive, lynch1991analysis}.  
However, subchondral bone remodelling in OA is not uniform.
In OA, the bone closest to the articular cartilage experiences the greatest effect and damage.
Therefore, alterations in subchondral bone texture are observed best at those regions and particularly at medial tibial compartment \cite{bayramoglu2019adaptive, mackay2017subchondral, janvier2015roi}.
The rest of the bone do not reveal texture differences in radiographs in such detail \cite{bayramoglu2019adaptive}.
Inspired by this phenomenon, rather than using the whole joint radiography we employed only the most informative region in our study (Figure \ref{fig:proposal}).
Here, we utilized again the landmark points to locate our ROI at medial tibia margin.
We used a  square  patch  with dimensions proportional to the width of the knee.

Previously, subchondral bone texture  in  the  radiographs  have  been   quantified mostly by fractal methods \cite{hirvasniemi2014quantification} and other conventional texture descriptors such as Local Binary Patterns \cite{bayramoglu2019adaptive}.
This is the first study to evaluate OA texture using a CNN model. 
We extracted texture features using a simple three layers CNN model (see Table \ref{tab:table1}).
The model is presented in Figure \ref{fig:proposal} (lower branch).
Considering the heavy
parameters of the current CNN-based methods in OA research that often use VGG or
ResNets family as backbones \cite{antony2016quantifying,tiulpin2018automatic,norman2019applying}, training from scratch is pretty difficult with limited training samples.
Thanks to the lightweight architecture of our CNN model, it can be trained from scratch with a good convergence.

\begin{table}[t]
  \renewcommand\arraystretch{0.9}
  \begin{center}
    \caption{Architecture of our tiny CNN. Each convolution layer (stride =1, padding =1) is followed by Batch normalization (BN), max pooling ($2\times 2$) and ReLU. For classification, we used two fully connected (FC) layers with droupout=0.5. }
    \label{tab:table1}
    \begin{tabular}{c|c|c}
      \hline
      \textbf{Layer} & \textbf{Output size} & \textbf{Architecture} \\
      \hline
      Input &  $48 \times 48$ & \\
      \hline
     Conv-1 $+$ BN $+$ MaxPool $+$ ReLU & $24 \times 24$ & $[3 \times 3, 32] $\\
      \hline
      Conv-2 $+$ BN $+$ MaxPool $+$ ReLU & $12 \times 12$ & $[3 \times 3, 64] $\\
      \hline
      Conv-3 $+$ BN $+$ MaxPool $+$ ReLU & $6 \times 6$ & $[3 \times 3, 128] $\\
      \hline
    \end{tabular}
  \end{center}
\end{table}

\section{Experiments}

We used the data  from the Osteoarthritis Initiative (OAI, \href{http://www.oai.ucsf.edu/ }{http://www.oai.ucsf.edu/}) and Multicenter Osteoarthritis Study (MOST, \href{http://most.ucsf.edu}{http://most.ucsf.edu}).
We selected all knees at baseline that are graded by KL score on both knees.
Missing entries were imputed and knees with total knee replacement were excluded.
The details about OAI and MOST datasets used in this study are presented in Table \ref{tab: data}.
As a strength of our paper, we validated our trained model with an independent test set (MOST) that is completely different from the training set (OAI) similar to \cite{tiulpin2018automatic, bayramoglu2019adaptive}. 

\begin{table}[!bt]
\caption{Description of the data. 
We used data from the OAI dataset and MOST dataset at baseline with postero-anterior view with 10 degrees beam angle (PA10) of the knee given that both joints are graded with KL score.
Missing entries were imputed and knees with total knee replacement were excluded.
}
\label{tab: data}
\centering
\begin{tabular}{lcc}
\toprule
 & \textbf{OAI} & \textbf{MOST} \\
  & Train Data & Test Data \\
\midrule

Number of knees & 8953  & 3445\\
Number of subjects & 4506 & 1748\\
Number of knees where KL$<$2, (non-OA)& 5045  & 2118\\
Number of knees where KL$\geq$2, (OA)& 3908  & 1327\\
\bottomrule
\end{tabular}
\end{table}

Firstly, landmark points ({keypoints}) of knee radiographs were extracted using an open software BoneFinder\textsuperscript{\textregistered} \cite{lindner2013fully}.
In the preprocessing pipeline, the 16-bit DICOM images are first normalized using global contrast normalisation and histogram truncation between the $5^{th}$ and $99^{th}$ percentiles and then converted to 8-bit images.
The resolution, which is not initially standardized in the database, is also standardized to $0.2$ mm.
Finally, using landmark points, each knee is rotated in order to have an aligned horizontal tibial plateau.
Subsequently, for texture characterization, we cropped a square patch having a side length of $1/7^{th}$ of the knee tibia width. 
Before feeding this texture ROI to our tiny CNN, it was rescaled to $56\times56$ pixels and random $48\times48$ pixel crops were then employed.

\begin{figure*}[!tb]
\centerline{
\subfloat[Feature Importance of $JS^2$]{\includegraphics[ width = 0.8\textwidth]{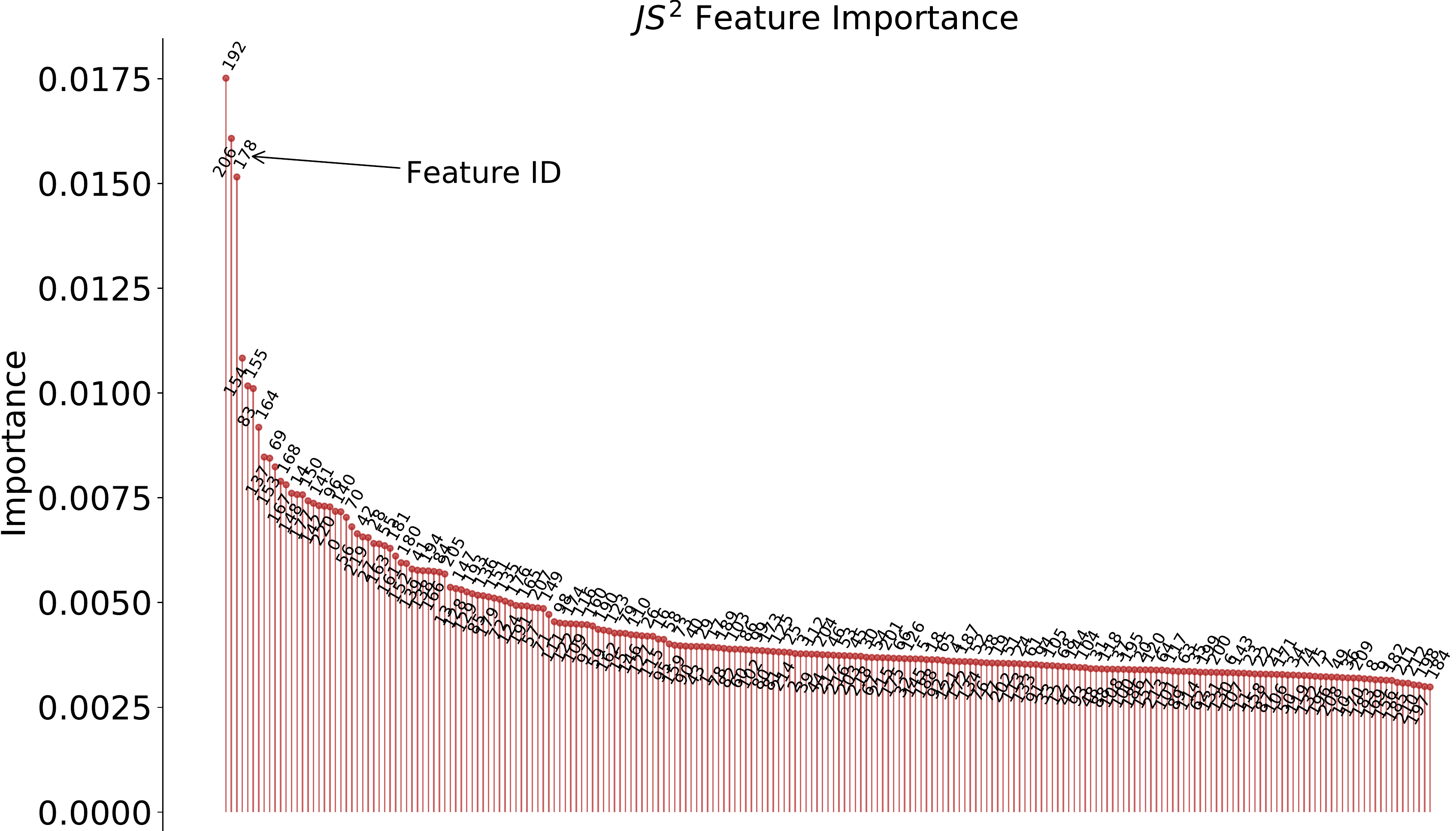}}
}
\centerline{
\subfloat[]{\includegraphics[height=5.5cm]{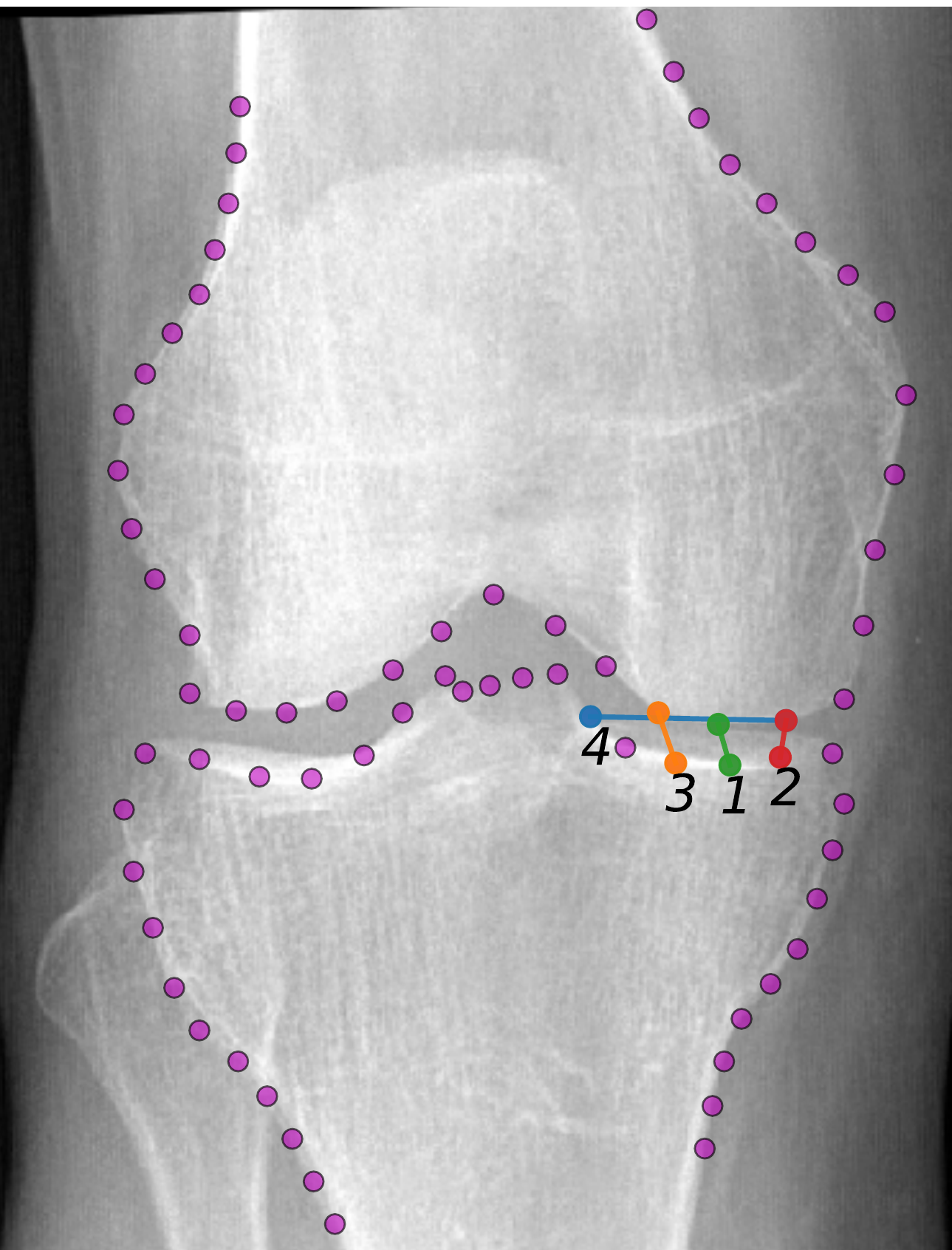}}
     }
\caption{Figure (a) shows the feature importance of $JS^2$ and (b) illustrates sample knee joint radiograph with landmarks and the most important four features of our $JS^2$ descriptor. }
\label{fig:jsw_imp}
\end{figure*}


\subsection{Reference Methods}
We used several reference methods to compare with the proposed approach.
For texture analysis, we employed LBP \cite{ojala2000gray}, Fractal Dimension (FD) \cite{lynch1991robust, hirvasniemi2016correlation}, Bilinear CNN (B-CNN) \cite{lin2017improved} and anti-alised CNN \cite{zhang2019making}.
Modern CNNs are not shift invariant and drastic changes at the network’s output can be observed due to one pixel
shift of an input image.
One possible solution is to use the antialiased networks \cite{zhang2019making} which utilizes signal processing technique (low-pass filtering before downsampling) to fix this.
We also tested B-CNN that captures pairwise correlations between the feature channels and showed promising results in fine-grained (texture) recognition problems \cite{lin2017improved}.

For shape-based analysis, we compared minJSW and fJSW which are commonly used in OA research \cite{duryea2003new, neumann2009location}.
Locations of fJSW measurements are illustrated in Figure \ref{fig:jsw} in which the knee joint is first aligned with respect to femur (i.e. x-axis is aligned with femoral condyles in both compartments).
The y-axis of the coordinate system is defined as the peripheral edge of the femoral condyle.
Subsequently, the fixed locations are selected at $x = 0.225$ (medialfJSW) and $x=0.8$ (lateralfJSW) following the literature \cite{duryea2003new, neumann2009location}. 
These measurements are  adjusted with respect to the tibia width.
Reference models based on LBP, FD, minJSW and fJSW  were assessed using Logistic Regression (LR) with `$L_2$' regularization using scikit-learn package \cite{scikit-learn}.

\subsection{Model, Training and Parameter Settings}

Our CNN model consists of 3 convolutional layers dedicated to texture feature extraction.
Each  convolution layer (stride=1, padding =1) is followed by Batch normalization (BN), max pooling ($2\times2$) and ReLU.
In our CNN-based experiments, we used two fully connected layers to make the prediction.
A dropout of 0.5 is inserted after the first fully connected layer.
For the combined model, where we concatenated shape and texture features, we adopted dropout of 0.3 as it provides better performance.

In all the experiments, we used the same training strategy.
We trained the models from scratch (end-to-end) using the random weight initialization.
We adopted stochastic gradient descent training on a GPU. 
A mini-batch of 64 images were employed, and a momentum of 0.9 was used and trained without weight decay. 
We used a starting learning rate of 0.01 and decreased it by 10 every 8 epochs. 
The models were trained for 100 epochs and we selected the model with a best performance.
Therefore, our results present an upper bound (Table \ref{tab: results}).
We trained our tiny CNN model also with data augmentations where we used small random rotations, gamma and brightness corrections.
However, it did not improve the scores.
This could be due to the knee alignment and pre-processing, which we performed before the ROI extraction.

For assessing the $JS^2$ descriptor we used both LR and two layer neural network (NN) which is similar to the last fully connected layers in the tiny CNN model.
We used PyTorch v1.4 in our experiments \cite{paszke2019pytorch}.

\subsection{Results}

In Table \ref{tab: results}, we present the area under the receiver operating characteristic curves (ROC AUC) to measure classifier performance effectively.

\begin{table}[t]
\caption{Comparison of the models' performance in detecting radiographic OA (KL$\geq$2). Reference models were assessed using Logistic Regression (LR). Here, Area under the ROC curve (AUC)
metric is used for assessment. NN: Neural Network. }
\label{tab: results}
\begin{center}
\noindent \resizebox{0.85\textwidth}{!}
{%

\begin{tabular}{llc}
\toprule
 & \textbf{Method} & \textbf{ROC AUC ({$\%$})} \\
\midrule

\multirow{5}{*}{\textbf{Texture}} & LBP (LR)& 82.69\\
& Fractal Dimension (LR)& 74.80 \\
& CNN & 88.74\\
& B-CNN \cite{lin2017improved}& 88.53\\
& Anti-alised CNN \cite{zhang2019making}& 87.93\\
\midrule 
\multirow{4}{*}{\textbf{Morphology}} & minJSW (LR)& 70.74\\
& minJSW + medfJSW + latfJSW (LR)& 81.75\\
& $JS^2$ Descriptor (LR) (ours) & 91.18\\
& $JS^2$ Descriptor (NN) (ours) & 93.52\\
& \begin{tabular}{@{}l@{}}$JS^2$ Descriptor  \\   \hspace{0.2cm}+ minJSW + medfJSW + latfJSW (NN)\end{tabular} & 93.60\\
\midrule
\textbf{Combined}& $JS^2$ Descriptor + CNN (ours) & \textbf{95.21} \\

\bottomrule
\end{tabular}

}
\end{center}
\end{table}

\textbf{Texture characterization.} We found that compared to the conventional methods, CNN-based texture extraction leads to better performance in detecting radiographic OA.
However, B-CNN and anti-alised CNN models did not improve the performance further.
Compared to the hand-crafted features, a plain tiny CNN model outperforms the best reference method by $6\%$.

\begin{SCfigure}
\centering

\includegraphics[width=0.55\textwidth]{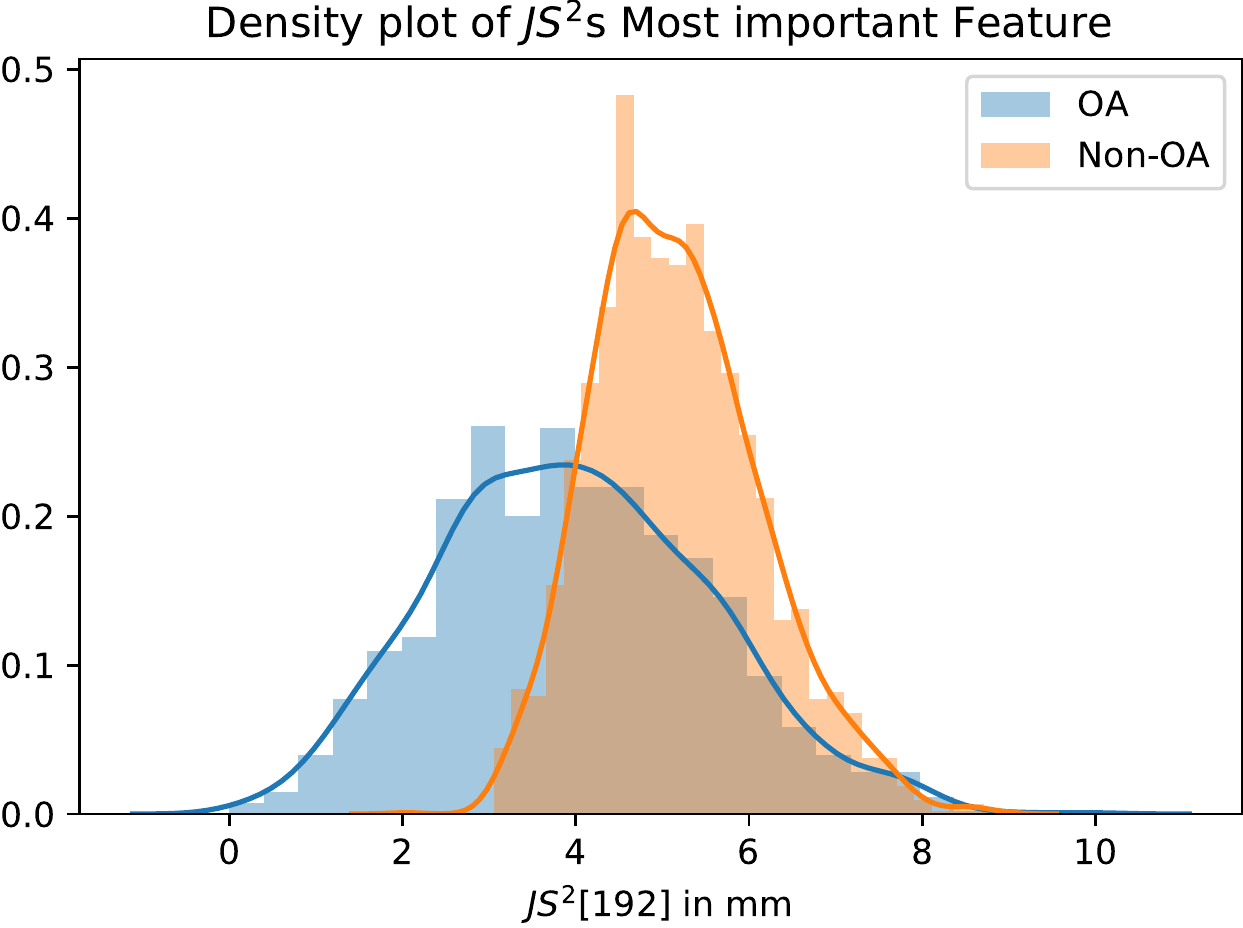}

\caption{\protect\rule{0ex}{8ex}Figure shows the density plot of $JS^2[192]$ feature which is the most important element in the descriptor. }
\label{fig:jsw192}
 \end{SCfigure}

\textbf{Joint Space - Joint Shape.} In this study, we evaluated minJSW for incident OA detection.
Then we combined minJSW measurement with JSW  measurements  at  fixed  locations  (fJSW), which improved the performance roughly $11\%$. 
The ROC AUC value achieved by our $JS^2$ method was $93.52\%$ which is significantly better than other approaches. 
In addition to the classifier performances, we investigated the feature importance of $JS^2$ (Figure \ref{fig:jsw_imp}).
Feature importance was calculated based on forests of trees classifier using scikit-learn package \cite{scikit-learn}.
Figure \ref{fig:jsw_imp}a illustrates how much each feature contributes to the model's overall predictive performance.
Most of the features are equally informative and, therefore, they cannot be ignored.
Figure \ref{fig:jsw_imp}b illustrates the first four most important elements of $JS^2$.
Interestingly, the fourth one might be related to knee alignment.
We evaluated the distribution of the most important $JS^2$ feature. 
Figure \ref{fig:jsw192} demonstrates the density plot of the most important feature for OA and and non-OA cases.
The mean (std) value of $JS^{2}[192]$ is $3.98mm(std\: 1.57mm)$ for OA and $5.17mm(std\: 0.96mm)$ for non-OA. 
Although not presented here, other features demonstrate similar characteristics.
Finally, we evaluated the robustness of our descriptor to JSW measurement errors and variability. 
The sources of errors and variability could be due to positioning accuracy in the acquisition of clinical knee X-rays and the landmark extraction method.
In order to simulate this, we added Gaussian distributed noise to our $JS^2$ descriptor with zero mean and standard deviation of 1mm, 3mm and 5mm to both train and test data.
Table \ref{tab:noise} demonstrates that our model is not sensitive to noisy JSW measurements.

To evaluate the true additive value of our $JS^2$ descriptor in the combined model, we assessed the descriptor's predictive ability both with LR and NN (Table \ref{tab: results}). 

\begin{table}[!b]
\begin{center}
\caption{Analysis of detection performance in case of noisy $JS^2$ measurements. Gaussian noise with zero mean is added to the $JS^2$ vector.}
\label{tab:noise}
\begin{tabular}{@{}lccc@{}}
\toprule
                               & \multicolumn{3}{c}{\textbf{ROC AUC (\%)}}                                                                                                                                                                \\ \midrule
   \textbf{Model}    & \textbf{\begin{tabular}[c]{@{}c@{}}Noise Std\\ 1mm\end{tabular}} & \textbf{\begin{tabular}[c]{@{}c@{}}Noise Std\\ 3mm\end{tabular}} & \textbf{\begin{tabular}[c]{@{}c@{}}Noise Std\\ 5mm\end{tabular}} \\ \midrule
$\mathbf{JS^2}$ & {93.21}   & {92.51}   & {91.81}       \\
\textbf{Combined ($\mathbf{JS^2}$ + CNN)}  & {94.85}     & {94.58}    & {94.17}  \\ \bottomrule
\end{tabular}
\end{center}
\end{table}

\textbf{Combined model}
As mentioned previously, we combined CNN-based texture features and our $JS^2$ descriptor to form a single vector and trained our classifier on fused feature (Figure \ref{fig:proposal}).
This model performed better than both texture and shape based models (ROC AUC = $95.21\%$).
This finding suggests that bone texture and joint shape carry complementary information for automatic detection of radiographic OA. 
The closest work to ours is Thomson et al. \cite{thomson2015automated} where they also combined shape and texture descriptors to detect OA. They reported $85\%$  ROC AUC using direct pixel ratios as texture descriptor and shape parameter vectors obtained from statistical shape modelling.
Moreover, our combined model achieves classification performance comparable, if not better, than the SOTA DL-based OA detection models \cite{tiulpin2018automatic, antony2016quantifying}.

\section{Discussions and Conclusions}

In this study, we propose a simple knee joint shape descriptor $JS^2$ for OA detection from plain radiographs.
In addition, we demonstrated that a lightweight CNN for extracting bone texture features together with our proposed joint shape-joint space descriptor achieves SOTA performance in radioraphic OA detection. 
Moreover, we evaluated the most important $JS^2$ features that contribute to OA presence. 
Unlike existing black-box deep CNN based models in the field, our descriptor provides more explainable results that translate features to an important structural change in joint anatomy that may contribute to OA.
We also demonstrated that our proposed $JS^2$ descriptor is robust to noisy joint space width measurements.

In our tiny CNN model, we utilized the most informative ROI (a very small image patch which is cropped from the medial tibia margin) to build a network with fewer parameters. 
Compared to the heavy deep CNN models that use  whole joint image, our tiny model still has the ability to recognize OA with high accuracy.
Our CNN model yields the highest score among the all previously published texture-based methods \cite{bayramoglu2019adaptive}.

This study has also some limitations.
First, our proposed descriptor is based on landmarks that needs to be extracted beforehand and this brings a preprocessing overhead.
However, similar overhead is also present in SOTA DL approaches for detecting the joint ROI from the full limb radiographs. 
Second, although medial OA is more common, other bone compartments (such as lateral and femur margins) which might hold new/complementary information of OA were not evaluated in our study.
Their inclusion could further improve the performance, which could be further investigated.
Admittedly, the models which utilize whole joint radiographs have the potential to utilize those features as a rule. 
Therefore, they are potentially better suited for fine-grained KL classification (KL 0, 1, 2, 3, and 4).

This study provides two important topics for future investigations. Firstly, fusing the proposed shape descriptor with the existing deep learning models to evaluate whether they capture shape-related features inherently or their performance can be improved by the explicit inclusion of shape features. 
Secondly, predicting the progression of OA using $JS^2$ descriptor.
Apart from OA, we believe that $JS^2$ descriptor could be useful in other fields where joint space features are important in disease assessment such as rheumatoid arthritis. 
Our source codes are available at \href{https://github.com/MIPT-Oulu/JS2}{https://github.com/MIPT-Oulu/JS2}.

\section*{Acknowledgments}

The OAI is a public-private partnership comprised of five contracts (N01-AR-2-2258; N01-AR-2-2259; N01-AR-2-2260; N01-AR-2-2261; N01-AR-2-2262)
funded by the National Institutes of Health, a branch of the Department of Health and Human Services, and conducted by the OAI Study Investigators. 
Private funding partners include Merck Research Laboratories; Novartis Pharmaceuticals Corporation, GlaxoSmithKline; and Pfizer, Inc. Private sector funding for the OAI is managed by the Foundation for the National Institutes of Health. This manuscript was prepared using an OAI public use data set and does not necessarily reflect the opinions or views of the OAI investigators, the NIH, or the private funding partners

Multicenter Osteoarthritis Study (MOST) Funding Acknowledgment. MOST is comprised of four cooperative grants (Felson – AG18820; Torner – AG18832, Lewis – AG18947, and Nevitt – AG19069) funded by the National Institutes of Health, a branch
of the Department of Health and Human Services, and conducted by MOST study investigators. This manuscript was prepared using MOST data and does not necessarily reflect the opinions or views of MOST investigators.

We would like to acknowledge the strategic funding of the University of Oulu, Infotech Oulu.

We gratefully  acknowledge the help received from Aleksei Tiulpin who extracted the landmarks using BoneFinder\textsuperscript{\textregistered} and the support of NVIDIA Corporation with the donation of the Quadro P6000 GPU used in this research.


\bibliographystyle{ieeetr}
\bibliography{bibliography.bib}

\begin{thebibliography}{10}

\bibitem{bijlsma2011osteoarthritis}
J.~W. Bijlsma, F.~Berenbaum, and F.~P. Lafeber, ``Osteoarthritis: an update
  with relevance for clinical practice,'' {\em The Lancet}, vol.~377, no.~9783,
  pp.~2115--2126, 2011.

\bibitem{altman2010early}
R.~D. Altman, ``Early management of osteoarthritis.,'' {\em The American
  journal of managed care}, vol.~16, pp.~S41--7, 2010.

\bibitem{heidari2011knee}
B.~Heidari, ``Knee osteoarthritis prevalence, risk factors, pathogenesis and
  features: Part i,'' {\em Caspian journal of internal medicine}, vol.~2,
  no.~2, p.~205, 2011.

\bibitem{Haq377}
I.~Haq, E.~Murphy, and J.~Dacre, ``Osteoarthritis,'' {\em Postgraduate Medical
  Journal}, vol.~79, no.~933, pp.~377--383, 2003.

\bibitem{kellgren1957radiological}
J.~Kellgren and J.~Lawrence, ``Radiological assessment of osteo-arthrosis,''
  {\em Annals of the rheumatic diseases}, vol.~16, no.~4, p.~494, 1957.

\bibitem{kose2018inter}
{\"O}.~K{\"o}se, B.~Acar, F.~{\c{C}}ay, B.~Yilmaz, F.~G{\"u}ler, and H.~Y.
  Y{\"u}ksel, ``Inter-and intraobserver reliabilities of four different
  radiographic grading scales of osteoarthritis of the knee joint,'' {\em The
  journal of knee surgery}, vol.~3, no.~03, pp.~247--253, 2018.

\bibitem{lynch1991analysis}
J.~Lynch, D.~Hawkes, and J.~Buckland-Wright, ``Analysis of texture in
  macroradiographs of osteoarthritic knees, using the fractal signature,'' {\em
  Physics in Medicine \& Biology}, vol.~36, no.~6, p.~709, 1991.

\bibitem{kraus2018predictive}
V.~B. Kraus, J.~E. Collins, H.~C. Charles, C.~F. Pieper, L.~Whitley, E.~Losina,
  M.~Nevitt, S.~Hoffmann, F.~Roemer, A.~Guermazi, {\em et~al.}, ``Predictive
  validity of radiographic trabecular bone texture in knee osteoarthritis: the
  osteoarthritis research society international/foundation for the national
  institutes of health osteoarthritis biomarkers consortium,'' {\em Arthritis
  \& Rheumatology}, vol.~70, no.~1, pp.~80--87, 2018.

\bibitem{hirvasniemi2014quantification}
J.~Hirvasniemi, J.~Thevenot, V.~Immonen, T.~Liikavainio, P.~Pulkkinen,
  T.~J{\"a}ms{\"a}, J.~Arokoski, and S.~Saarakkala, ``Quantification of
  differences in bone texture from plain radiographs in knees with and without
  osteoarthritis,'' {\em Osteoarthritis and cartilage}, vol.~22, no.~10,
  pp.~1724--1731, 2014.

\bibitem{thomson2015automated}
J.~Thomson, T.~O’Neill, D.~Felson, and T.~Cootes, ``Automated shape and
  texture analysis for detection of osteoarthritis from radiographs of the
  knee,'' in {\em International Conference on Medical Image Computing and
  Computer-Assisted Intervention}, pp.~127--134, Springer, 2015.

\bibitem{janvier2017subchondral}
T.~Janvier, R.~Jennane, H.~Toumi, and E.~Lespessailles, ``Subchondral tibial
  bone texture predicts the incidence of radiographic knee osteoarthritis: data
  from the osteoarthritis initiative,'' {\em Osteoarthritis and cartilage},
  vol.~25, no.~12, pp.~2047--2054, 2017.

\bibitem{antony2016quantifying}
J.~Antony, K.~McGuinness, N.~E. O'Connor, and K.~Moran, ``Quantifying
  radiographic knee osteoarthritis severity using deep convolutional neural
  networks,'' in {\em Pattern Recognition (ICPR), 2016 23rd International
  Conference on}, pp.~1195--1200, IEEE, 2016.

\bibitem{norman2019applying}
B.~Norman, V.~Pedoia, A.~Noworolski, T.~M. Link, and S.~Majumdar, ``Applying
  densely connected convolutional neural networks for staging osteoarthritis
  severity from plain radiographs,'' {\em Journal of digital imaging}, vol.~32,
  no.~3, pp.~471--477, 2019.

\bibitem{tiulpin2018automatic}
A.~Tiulpin, J.~Thevenot, E.~Rahtu, P.~Lehenkari, and S.~Saarakkala, ``Automatic
  knee osteoarthritis diagnosis from plain radiographs: a deep learning-based
  approach,'' {\em Scientific reports}, vol.~8, no.~1, p.~1727, 2018.

\bibitem{gorriz2019assessing}
M.~G{\'o}rriz, J.~Antony, K.~McGuinness, X.~Gir{\'o}-i Nieto, and N.~E.
  O’Connor, ``Assessing knee oa severity with cnn attention-based end-to-end
  architectures,'' in {\em International Conference on Medical Imaging with
  Deep Learning}, pp.~197--214, 2019.

\bibitem{abedin2019predicting}
J.~Abedin, J.~Antony, K.~McGuinness, K.~Moran, N.~E. O’Connor,
  D.~Rebholz-Schuhmann, and J.~Newell, ``Predicting knee osteoarthritis
  severity: comparative modeling based on patient’s data and plain x-ray
  images,'' {\em Scientific reports}, vol.~9, no.~1, pp.~1--11, 2019.

\bibitem{minciullo2016fully}
L.~Minciullo and T.~Cootes, ``Fully automated shape analysis for detection of
  osteoarthritis from lateral knee radiographs,'' in {\em Pattern Recognition
  (ICPR), 2016 23rd International Conference on}, pp.~3787--3791, IEEE, 2016.

\bibitem{haverkamp2011variation}
D.~J. Haverkamp, D.~Schiphof, S.~M. Bierma-Zeinstra, H.~Weinans, and J.~H.
  Waarsing, ``Variation in joint shape of osteoarthritic knees,'' {\em
  Arthritis \& Rheumatism}, vol.~63, no.~11, pp.~3401--3407, 2011.

\bibitem{martinez2019discovering}
A.~M. Martinez, I.~Flament, F.~Liu, J.~Lee, P.~Cao, S.~Majumdar, and V.~Pedoia,
  ``Discovering knee osteoarthritis bone shape features using deep learning,''
  {\em Osteoarthritis and Cartilage}, vol.~27, pp.~S386--S387, 2019.

\bibitem{ahlback1968osteoarthrosis}
S.~Ahlback, ``Osteoarthrosis of the knee. a radiographic investigation,'' {\em
  Acta radiol.}, vol.~227, pp.~7--72, 1968.

\bibitem{geirhos2018imagenet}
R.~Geirhos, P.~Rubisch, C.~Michaelis, M.~Bethge, F.~A. Wichmann, and
  W.~Brendel, ``Imagenet-trained cnns are biased towards texture; increasing
  shape bias improves accuracy and robustness,'' {\em arXiv preprint
  arXiv:1811.12231}, 2018.

\bibitem{bayramoglu2019adaptive}
N.~Bayramoglu, A.~Tiulpin, J.~Hirvasniemi, M.~T. Nieminen, and S.~Saarakkala,
  ``Adaptive segmentation of knee radiographs for selecting the optimal roi in
  texture analysis,'' {\em Osteoarthritis and Cartilage}, 2020.

\bibitem{buckland2004subchondral}
C.~Buckland-Wright, ``Subchondral bone changes in hand and knee osteoarthritis
  detected by radiography,'' {\em Osteoarthritis and cartilage}, vol.~12,
  pp.~10--19, 2004.

\bibitem{kamibayashi1995trabecular}
L.~Kamibayashi, U.~Wyss, T.~Cooke, and B.~Zee, ``Trabecular microstructure in
  the medial condyle of the proximal tibia of patients with knee
  osteoarthritis,'' {\em Bone}, vol.~17, no.~1, pp.~27--35, 1995.

\bibitem{goldring2009role}
S.~R. Goldring, ``Role of bone in osteoarthritis pathogenesis,'' {\em Medical
  Clinics of North America}, vol.~93, no.~1, pp.~25--35, 2009.

\bibitem{mehta2017comparison}
N.~Mehta, J.~Duryea, G.~J. Badger, M.~R. Akelman, M.~H. Jones, K.~P. Spindler,
  and B.~C. Fleming, ``Comparison of 2 radiographic techniques for measurement
  of tibiofemoral joint space width,'' {\em Orthopaedic journal of sports
  medicine}, vol.~5, no.~9, p.~2325967117728675, 2017.

\bibitem{neumann2009location}
G.~Neumann, D.~Hunter, M.~Nevitt, L.~Chibnik, K.~Kwoh, H.~Chen, T.~Harris,
  S.~Satterfield, J.~Duryea, {\em et~al.}, ``Location specific radiographic
  joint space width for osteoarthritis progression,'' {\em Osteoarthritis and
  cartilage}, vol.~17, no.~6, pp.~761--765, 2009.

\bibitem{guermazi2009plain}
A.~Guermazi, D.~J. Hunter, and F.~W. Roemer, ``Plain radiography and magnetic
  resonance imaging diagnostics in osteoarthritis: validated staging and
  scoring,'' {\em JBJS}, vol.~91, no.~Supplement\_1, pp.~54--62, 2009.

\bibitem{duryea2003new}
J.~Duryea, S.~Zaim, and H.~Genant, ``New radiographic-based surrogate outcome
  measures for osteoarthritis of the knee,'' {\em Osteoarthritis and
  cartilage}, vol.~11, no.~2, pp.~102--110, 2003.

\bibitem{lindner2013fully}
C.~Lindner, S.~Thiagarajah, J.~M. Wilkinson, G.~A. Wallis, T.~F. Cootes,
  arcOGEN Consortium, {\em et~al.}, ``Fully automatic segmentation of the
  proximal femur using random forest regression voting,'' {\em IEEE
  transactions on medical imaging}, vol.~32, no.~8, pp.~1462--1472, 2013.

\bibitem{aspden2019osteoarthritis}
R.~M. Aspden and F.~Saunders, ``Osteoarthritis as an organ disease: from the
  cradle to the grave,'' {\em European cells \& materials}, 2019.

\bibitem{mackay2017subchondral}
J.~W. MacKay, P.~J. Murray, B.~Kasmai, G.~Johnson, S.~T. Donell, and A.~P.
  Toms, ``Subchondral bone in osteoarthritis: association between mri texture
  analysis and histomorphometry,'' {\em Osteoarthritis and cartilage}, vol.~25,
  no.~5, pp.~700--707, 2017.

\bibitem{janvier2015roi}
T.~Janvier, H.~Toumi, K.~Harrar, E.~Lespessailles, and R.~Jennane, ``Roi impact
  on the characterization of knee osteoarthritis using fractal analysis,'' in
  {\em Image Processing Theory, Tools and Applications (IPTA), 2015
  International Conference on}, pp.~304--308, IEEE, 2015.

\bibitem{ojala2000gray}
T.~Ojala, M.~Pietik{\"a}inen, and T.~M{\"a}enp{\"a}{\"a}, ``Gray scale and
  rotation invariant texture classification with local binary patterns,'' in
  {\em European Conference on Computer Vision}, pp.~404--420, Springer, 2000.

\bibitem{lynch1991robust}
J.~Lynch, D.~Hawkes, and J.~Buckland-Wright, ``A robust and accurate method for
  calculating the fractal signature of texture in macroradiographs of
  osteoarthritic knees,'' {\em Medical Informatics}, vol.~16, no.~2,
  pp.~241--251, 1991.

\bibitem{hirvasniemi2016correlation}
J.~Hirvasniemi, J.~Thevenot, H.~T. Kokkonen, M.~A. Finnil{\"a}, M.~S.
  Ven{\"a}l{\"a}inen, T.~J{\"a}ms{\"a}, R.~K. Korhonen, J.~T{\"o}yr{\"a}s, and
  S.~Saarakkala, ``Correlation of subchondral bone density and structure from
  plain radiographs with micro computed tomography ex vivo,'' {\em Annals of
  biomedical engineering}, vol.~44, no.~5, pp.~1698--1709, 2016.

\bibitem{lin2017improved}
T.-Y. Lin and S.~Maji, ``{Improved Bilinear Pooling with CNNs},'' in {\em
  BMVC}, 2017.

\bibitem{zhang2019making}
R.~Zhang, ``Making convolutional networks shift-invariant again,'' {\em arXiv
  preprint arXiv:1904.11486}, 2019.

\bibitem{scikit-learn}
F.~Pedregosa, G.~Varoquaux, A.~Gramfort, V.~Michel, B.~Thirion, O.~Grisel,
  M.~Blondel, P.~Prettenhofer, R.~Weiss, V.~Dubourg, J.~Vanderplas, A.~Passos,
  D.~Cournapeau, M.~Brucher, M.~Perrot, and E.~Duchesnay, ``Scikit-learn:
  Machine learning in {P}ython,'' {\em Journal of Machine Learning Research},
  vol.~12, pp.~2825--2830, 2011.

\bibitem{paszke2019pytorch}
A.~Paszke, S.~Gross, F.~Massa, A.~Lerer, J.~Bradbury, G.~Chanan, T.~Killeen,
  Z.~Lin, N.~Gimelshein, L.~Antiga, {\em et~al.}, ``Pytorch: An imperative
  style, high-performance deep learning library,'' in {\em Advances in NIPS},
  pp.~8024--8035, 2019.

\end{thebibliography}

\end{document}